# Real-space study of the growth of magnesium on ruthenium

T. Herranz[1], B. Santos[1], K. F. McCarty[2], J. de la Figuera[1]
1 Instituto de Química-Física "Rocasolano", CSIC, Madrid 28006, Spain
2 Sandia National Laboratories, Livermore, California 94550, USA

Abstract
The growth of magnesium on ruthenium has been studied by low-energy electron microscopy (LEEM) and scanning tunneling microscopy (STM). In LEEM, a layer-by-layer growth is observed except in the first monolayer, where the completion of the first layer in inferred by a clear peak in electron reflectivity. Desorption from the films is readily observable at 400 K. Real-space STM and low-energy electron diffraction confirm that sub-monolayer coverage presents a moiré pattern with a 12 Å periodicity, which evolves with further Mg deposition by compressing the Mg layer to a 22 Å periodicity. Layer-by-layer growth is followed in LEEM up to 10 ML. On films several ML thick a substantial density of stacking faults are observed by dark-field imaging on large terraces of the substrate, while screw dislocations appear in the stepped areas. The latter are suggested to result from the mismatch in heights of the Mg and Ru steps. Quantum size effect oscillations in the reflected LEEM intensity are observed as a function of thickness, indicating an abrupt Mg/Ru interface.

## 1 Introduction

Magnesium is a commonly available non-toxic metal. From a technology point of view, its hydride $MgH_2$ has been proposed as a lightweight hydrogen carrier, even if kinetic and thermodynamic limitations have limited its use. From a fundamental point of view, Mg together with Be and Al have been studied as part of the so-called free-electron-like metals with quite ideal metallic bonding. The simple electronic structure of Mg and its nearly ideal hexagonal close-packed lattice, with a *c/a* lattice parameter ratio of 1.624 compared to the ideal value of 1.633, simplifies fundamental studies. However, its low sublimation temperature limits the preparation of high quality single crystals in ultra-high vacuum (UHV). Fortunately, Mg is known to grow as highly perfect thin films on many substrates. On some of them, the interface with the substrate is very sharp. With the appropriate substrate, quantization of the Mg sp-bands has been observed up to a thickness of several nanometers[1].

There are only a limited number of studies on the growth of Mg on refractory metals, most of them on W(110)[1, 2]. On W(110), a real-space study by low-energy electron microscopy[2] found a very high quality layer-by-layer growth. The quality of such thin films is so good that is has been used to study how the bulk electronic structure develops as a function of thickness[1] and influences chemical reactivity[3].

Over et al.[4] characterized the epitaxial growth of Mg over Ru(0001) using LEED, work function changes and temperature programmed desorption (TPD). They reported that magnesium grows over Ru(0001) in an incommensurate manner, i.e., keeping its own in-plane spacing. This was understood to be due to the large mismatch between the respective in-plane lattice spacings (around 18%). The result is an overlayer film without significant strain. A moiré was observed in the LEED pattern, starting at coverages as low as 0.05 Magnesium monolayers (ML). From $\theta$=0.05-0.65 ML the LEED indicated a (5x5) periodicity with a Mg-Mg distance of 3.35 Å (4% expansion with respect to the Mg bulk value). When the coverage was increased to 0.65-0.75 ML, a compression of the Mg overlayer was observed, with the LEED pattern changing gradually to a (7x7) periodicity with a final Mg-Mg distance of 3.13 Å (slightly compressed with respect to the bulk distance of 3.21 Å). When $\theta$=0.75 ML there were no further changes in the LEED patterns and the authors considered that the 1 layer of Mg/Ru completely covered the substrate (i.e., its coverage was 1 ML= 0.75 ML). From TPD measurements the authors described several peaks. For a coverage of less than one monolayer of Mg, there were three main TPD peaks: $\alpha$ at 750 K, which is observed for small Mg coverages, $\beta$ at 580 K, corresponding to the compressed phase (7x7) and $\gamma$ at 500 K, which is present for higher Mg coverages. This last peak shifts to lower temperatures when the coverage is higher than one monolayer. Specifically the authors assigned the peak $\gamma$ at 550 K to the desorption of the 2$^{nd}$ ML, the peak $\gamma'$ at 530 K to the desorption of the 3$^{rd}$ one and the peak $\gamma''$ at 510 K to the desorption to the 4$^{th}$ and subsequent layers. From these results it can be deduced that the first Mg layer interacts more strongly with the support than the subsequent layers interact with Mg layers. The same group published later[5] a LEED-IV fit providing a crystallographic structure of the first 3 ML of Mg over Ru(0001). They used the (5x5) symmetry for the analysis of the first Mg overlayer. The second and third overlayers were treated within a mirror approximation, giving the films a 7/6$x$7/6 structure.

A few works have described STM imaging of epitaxial Mg grown at room temperature up to 2 ML. Pezzegara et al. [6] described recently the growth of continuous Mg films over the semiconductor GaN(0001) using STM and reflection high-energy electron diffraction (RHEED). The in-plane lattice mismatch with Mg is only 0.3%. In agreement with this small mismatch, they did not observe a moiré pattern on the Mg surface. They reported hexagonal shaped Mg islands with a height of 2.80 Å (slight larger than the interlayer spacing of close-packed planes in bulk Mg, 2.60 Å). When the coverage was lower than 0.4 ML the islands displayed atoll-like shapes. Increasing the Mg coverage changed the shape of the islands, making their shape more compact. The appearance of the islands depended also on the STM imaging bias. With 4 V the islands looked flat. However, if the bias was lowered to 0.5 V there was a detectable corrugation at their surface of 0.06 Å(6 pm). Submonolayer Mg over Si(001) was characterized by Hutchinson et al. [7] at room temperatures using STM. Deposited magnesium formed rows that are roughly perpendicular to the substrate dimer rows.

In this paper we present a real-space STM and LEEM study of the growth of Mg films up to 10 ML on Ru(0001). Since this substrate does not alloy with Mg under growth conditions and both materials have the same crystal structure, the growth processes are simplified and the Mg films closer approach bulk material. The knowledge of how Mg grows over Ru(0001) is a key part of our group's efforts in understanding the relationship between the atomic structure of ultrathin films and their hydrogenation/dehydrogenation ability as a hydrogen storage material[8]. In the present work we obtain LEED data in a LEEM microscope that confirms the results obtained previously by Over and coworkers. New insight is obtained from present real space data of the Mg growing acquired using two complementary microscopic techniques (LEEM and STM). In thin films, STM reveals corrugations at the same periodicities as the moiré between the Mg and Ru lattices. In thicker films, we find stacking faults between Mg regions and screw dislocations in regions of high Ru step density.

## 2 Experimental details

The magnesium growth on ruthenium was performed in two different UHV chambers with two Ru(0001) single crystals as substrates. The first chamber has a commercial low-energy electron microscope (Elmitec III). The microscope can monitor growth in real time, or during heating/cooling of the substrate between 200-1600 K. The second houses a low-energy electron diffractometer and a home-made scanning tunneling microscope (STM)[9]. The STM is controlled by commercial RHK electronics and the open-source Gxsm STM software[10, 11]. For analysis, we used the packages Gwyddion[12] and ImageJ[13] for STM and LEEM images, respectively. The base pressure of both UHV systems is below $1\times10^{-10}$ mbar. In the LEEM system, the Ru(0001) substrate was cleaned by exposure to $1.5\times10^{-8}$ mbar of O at 890 K, followed by brief flashes to 1600 K. In the STM system, the substrate was cleaned by repeated exposure and flash cycles (exposure to 20 sec at $1.3\times10^{-7}$ mbar at room temperature and flashing to 1600 K). Mg was deposited from a Mg rod heated by electron bombardment. During the film growth the pressure remained in the low $10^{-10}$ mbar range. Typical deposition rates were 1 ML/min. A monolayer is defined as a bulk-like Mg layer [4].

The LEEM images directly the electrons reflected by the surface under observation. Thus, the measurement of the averaged reflected intensity, either as a function of deposition time, or as a function of the incoming electron energy, is extracted by averaging the image intensity from a suitable region (box). When imaging the diffraction pattern, the relevant data is the integrated intensity and position of each spot. When measuring the spatial position of a diffracted spot in LEED, a 2-dimensional Gaussian function was fitted to the spot position.

## 3 Results and Discussion

### 3.1 Growth in LEEM

The low-energy electron microscopy view of the first stages of Mg growth on Ru(0001) is shown in Figure 1. The LEEM snapshots are selected from a sequence acquired while continuously imaging the surface during the magnesium deposition. Growth of first-layer islands cannot be imaged directly by LEEM. At first, only a uniform decrease of the reflected electron intensity is detected and the substrate steps become less visible. In Figure 1g the spatially averaged reflected intensity is plotted as a function of the deposition time. After reaching a minimum, the reflected electron intensity increases again. Completion of the first layer is indicated by the maximum of the reflected intensity at 75 s. At the same time, in the real-space images, the substrate steps are suddenly visible again. Nevertheless the Mg-covered substrate terraces are not imaged like the bare Ru substrate. The fine-scale contrast in Figure 1b suggests that there are structures in the monolayer whose size is close to the resolution limit of the LEEM, which for our instrument is about 100 Å. The same observation (lack of first monolayer LEEM island contrast, with the substrate steps clearly visible at close to the compact monolayer) has been reported for Mg/W(110)[2] and ascribed to the formation of small islands.

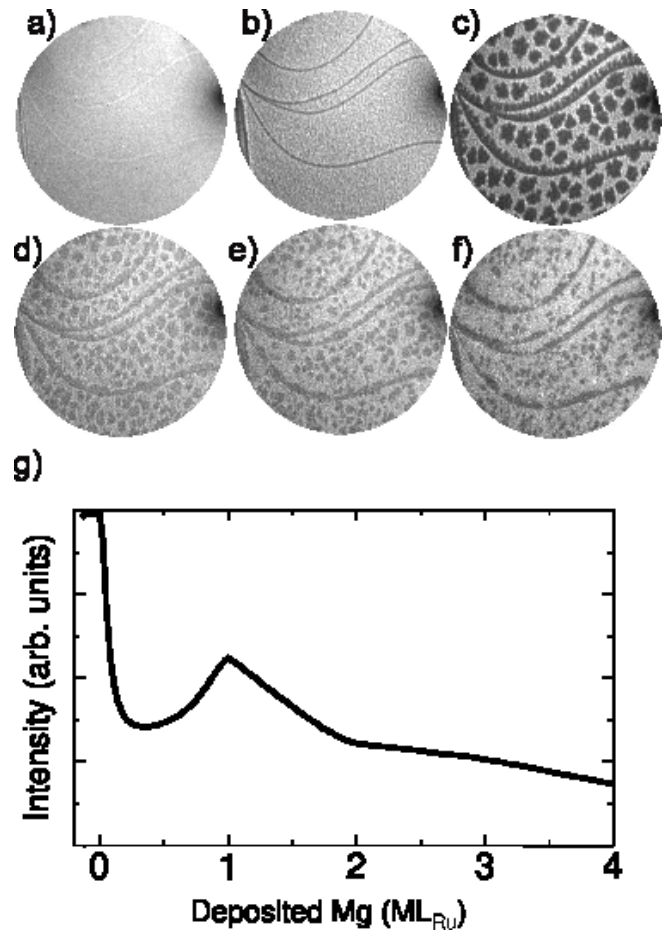

Figure 1: a)-f) LEEM snapshots from a sequence acquired while growing Mg on Ru(0001) at 373 K. The field of view (FOV) is 7 μm and the electron energy is 5 eV. a) shows the bare substrate. b) corresponds with the substrate just before the appearance of the 2 ML islands. c),d),e) and f) show islands 2, 3, 4, 5 ML thick respectively. The first ML islands cannot be distinguished in LEEM [b)], but the surface looks rough when compared with the Ru substrate [a)]. In g) the average reflected intensity is plotted versus time. The first peak corresponds to the completion of the first ML [b)].

The nucleation of the second layer, by contrast, is clearly observed, and proceeds through well-defined island growth. The nucleation density increases slightly for the next layers. At the electron energy of 5 eV, no clear oscillations of the reflected intensity are detected beyond the monolayer. The islands are roughly round in shape, as expected for a three fold structure on the three fold symmetric Ru(0001) substrate. (In contrast, there is a strong uniaxial island growth in Mg/W(110) due to the anisotropy induced by the W(110) substrate.)

In Figure 2, a growth sequence is shown up to a thickness close to 10 ML. The electron energy was varied to achieve optimum interlayer contrast for the different thickness. The growth is completely layer by layer, with no more than 3 different thicknesses exposed at a given time except at step bunches, where a more three-dimensional growth is observed. All together, the quality of the growth is very high, improving with thickness.

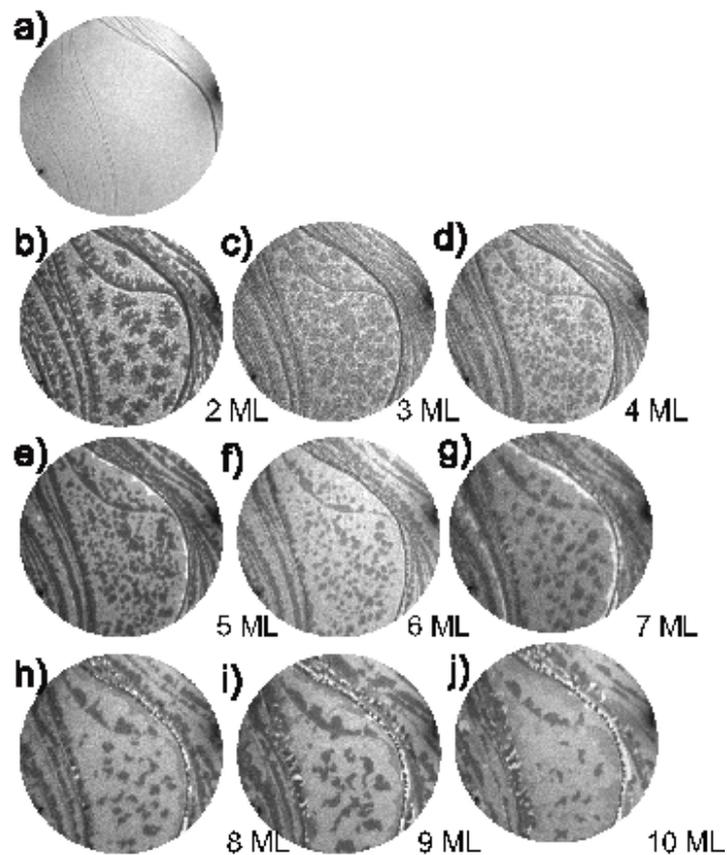

Figure 2: a)-j) Sequence of LEEM images showing the growth of Mg on Ru(0001) up to a thickness of 10 ML. The field of view is 7 μm. The electron energy has been varied between 3–7 eV to maximize contrast between consecutive layers. The growth temperature is 383 K. The exact thicknesses of the tallest islands at the center of the large substrate terrace are labeled.

In Figure 3a-c, the growth of the second layer is followed at the slightly higher temperature of 408 K. The islands have now a dendritic shape, very different from the compact shapes observed at lower temperatures. Dendritic islands have been observed for many metal/metal growth systems. In particular, wide-arm dendritic islands have been explained as resulting either from alloy formation changing surface diffusion[14] or to instabilities in the diffusion field around growing islands, as, for example, diffusion limited aggregation (DLA) of Au/Ru(0001) [15, 16]. In our case, it is striking that the islands are compact at a lower temperature, and dendritic at a higher temperature. This might indicate alloy formation. But the temperature is still extremely low for ruthenium to alloy with Mg, and the observed islands correspond to the nucleation of the second layer over a single continuous monolayer film. If alloying would happen, it would be expected to occur also at the monolayer islands. Furthermore if the deposition is stopped, the dendritic islands slowly disappear, as show in the Figure 3d-f and summarized in the island size evolution plotted in Figure 3g (a similar observation was reported for Mg on W(110)[2]). That the 2 ML islands disappearance is due to alloying is also highly unlikely. Reported work by TDS of multilayer Mg films on Ru[4] indicates that the third Mg layer is bound less strongly that the second. In agreement with this observation, we have been unable to nucleate any third layer islands at the conditions of the experiment shown in Fig. 3a. LEEM experiments (not shown) on the desorption of multilayer films grown at lower temperatures (such as the film presented in Fig. 2) also support this interpretation. 2

We thus propose that the observed dendritic island growth is due to a combination of energetics that includes significant sublimation. Actually, even while the islands are disappearing at the measurement temperature, there is hardly any change in the island shape, suggesting a larger edge-diffusion energy barrier, compared to the desorption energy barrier. That is, the edge adatoms of the second monolayer islands desorb from the surface before they have a chance of moving around the island, in marked contrast with most metal/metal systems.

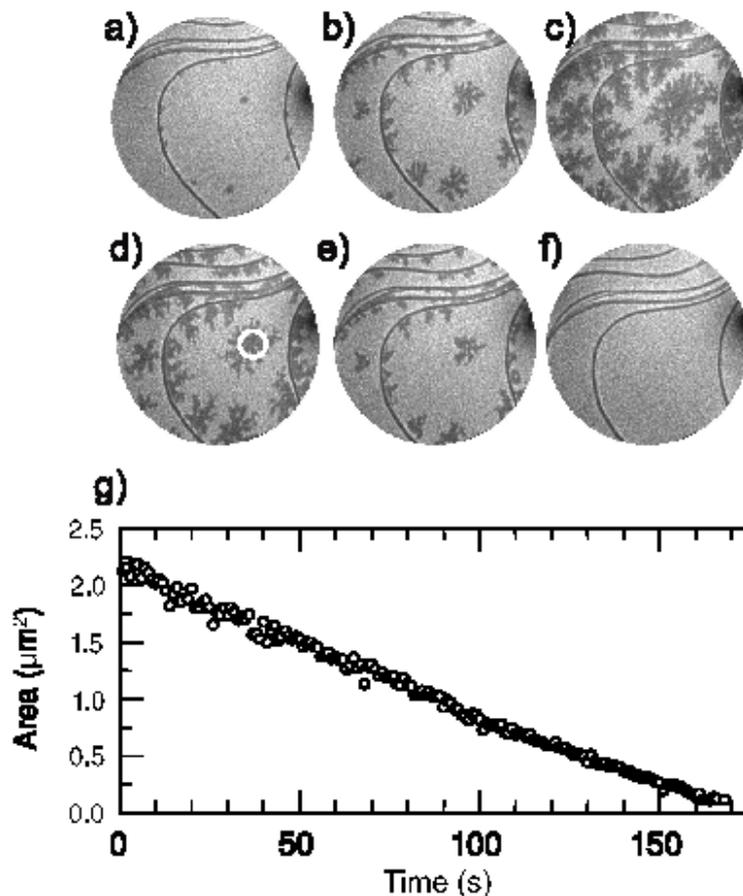

Figure 3: Growth and sublimation of Mg at a sample temperature of 408 K. Images a)-c) correspond to the deposition of Mg at a rate of 1 ML/minute. After frame c), the Mg flux was stopped. Frames d), e) and f) were acquired 99, 172 and 233 s after frame c). The field of view is 7μm, and the electron energy is 5 eV. g) Plot of the area of the island marked in d) with a white circle as a function time.

## 3.2 LEED in LEEM

Low-energy electron diffraction (LEED) patterns were obtained in the LEEM microscope[17, 18] by changing the power of the imaging lenses to display the back-focal plane of the objective lens. In Figure 4, the LEED pattern acquired from a 2 μm diameter-area of the sample is presented for the Ru(0001) substrate (Figure 4a), the initial pattern observed when there is a coverage below 0.65 (with respect to the substrate, Figure 4b), and the pattern observed when the 2 layer starts to nucleate

(Figure 4c). The images are snapshots from a sequence of LEED patterns acquired while growing Mg at a slower rate than the previous films, with the substrate kept at 357 K. The LEED patterns present sixfold symmetry because they were acquired from several substrate terraces. As the LEED pattern from adjacent terraces present rotated three-fold symmetric patterns, they average to the observed six-fold pattern[17]. Initially only the Ru LEED is detected, with the unreconstructed 1x1 periodicity of the hcp(0001) bare surface. At a coverage of 0.6 ML, additional spots start to appear, forming a superstructure of a periodicity close to (5x5) where only the first superstructure order spots are detected. The most straightforward interpretation of the LEED pattern is a coincidence pattern between the Mg film and the underlying Ru substrate[4], i.e., a moiré pattern on the Mg film. By calibrating the Mg spot separations using the Ru spots, we estimate the initial Mg in-plane lattice spacing to be 3.45±0.07Å. The 2% error arises from distortions in the imaging lenses of our LEEM, as estimated from comparing the distances measured using different, equivalent spots. The lattice spacing initially stays roughly constant before gradually decreasing to 3.10±0.06Å, as show in Figure 4(d). This periodicity is close to a 7x7 pattern. With more Mg deposition, the intensities of the Ru spots decrease until only a 1x1 pattern of Mg first-order beams remains. The lattice spacing decrease shown in Figure 4(d) occurs when the surface is already covered with a layer of magnesium. Further magnesium deposition densifies the already complete monolayer. Only when the first layer is completely dense do $2^{nd}$ layer islands nucleate. As discussed in the experimental details, we define 1 ML as the complete layer corresponding to the "compact" Mg/Ru phase. Our LEED/LEEM in-plane distance measurements are in agreement with the LEED results obtained with the standard LEED diffractometer used in Ref. [4]. In addition, in Figure 4(e), the evolution of the spot width with coverage in the same coverage range is presented. The spot width of a LEED beam can be related to the average domain or island size of the surface. In our case, the spot width of the Ru beams is quite constant (close to 1% of the spot distance to the specular beam). This width can be considered the error limit of our measurements (the LEED data was acquired on a single Ru substrate terrace, so there is no influence of substrate steps). On the other hand, the Mg spot starts with a larger size, and decreases with coverage from 3.5% to 1.7% even before the change in lattice spacing starts. This implies that the domain size increases during the transition. Using the estimate $h_s/a^* \sim a/d$ (where $h_s/a^*$ is the half-width of the Mg spot with respect to the Mg spot separation with the specular beam, and a is the atomic spacing f the Mg film and d is the domain size) the domain size increases from close to 100 Å before the compression begins to about 180 Å when the 7x7 pattern is reached. It is clear from the spot evolution that there is no coexistence of 7x7 and 5x5, but rather a continuous compression of the Mg layer.

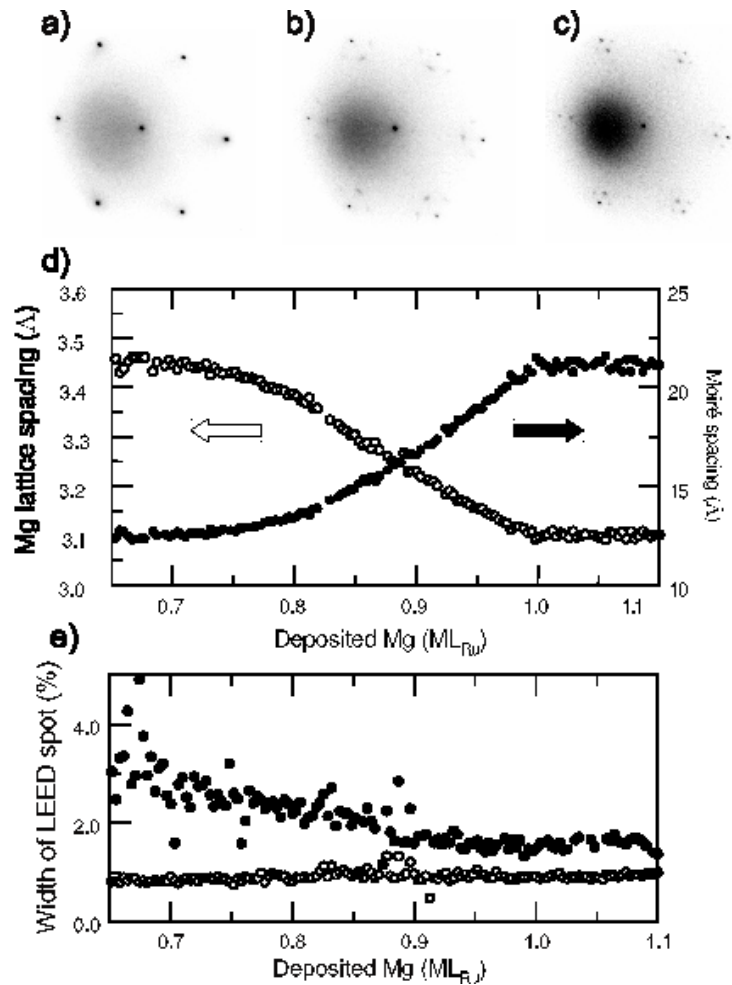

Figure 4: a)-c) LEED snapshots at an electron energy of 42 eV from a sequence acquired while growing 2 ML of Mg on Ru(0001). The substrate temperature is 357 K. a) Bare Ru. b) Initial pattern showing a periodicity close to 5x5 (coverage between 0.50 and 0.65) c) Final pattern with a periodicity of 7x7. Thicker coverages show the same pattern with weaker Ru spots until only the Mg spots can be seen. d) Evolution, measured from the sequence of LEED patterns of the real-space in-plane lattice spacing of the Mg beams (empty circles) and the distance in real space between the Mg and Ru beams (filled circles), i.e., the distance corresponding to the moiré periodicity. e) Evolution of the spot width of the Mg (filled circles) and the Ru (empty circles) first order beams, normalized by the each spot (Mg or Ru respectively) distance to the specular beam.

## 3.3 Reflectivity in LEEM

In Figure 5 we show how the energy-dependent electron reflectivity evolves with Mg thickness[18]. The data was extracted from a sequence of LEEM images acquired at a low deposition rate of 0.12 ML/min while, at the same time, scanning the electron energy in the range 3-30 eV at a rate of 68 s per energy scan. The reflected intensity is measured by averaging the image intensity within a box several micrometers in size. This acquisition procedure is usually employed when spatially resolved information is not available (such as in averaged reflectivity measurements [19, 20], or in LEEM studies of films where the growing islands cannot be resolved [21]) and mirrors the method used for the occupied electronic structure determined by valence band photoemission spectroscopy [22]. In our case, we use it to obtain both reflectivity curves of uniformly thick layers (see individual plots in Figure 5) as well as the intermediate thicknesses (see inset of Fig. 5).

In the plots of Fig. 5 we have separated the reflectivity of the thinner film (bottom panel) and the thicker one (top panel). The first scan shown corresponds to the bare Ru substrate, with a broad peak at the 0002 bulk Bragg reflection near 20 eV. As Mg is deposited, the electron reflectivity changes, and eventually, reaches the reflectivity of a Mg bulk-like surface. Although from a structural point of view even the first compact Mg layer is very close to a bulk-Mg surface, the electron reflectivity is very different from bulk Mg due to the interaction with the Ru substrate. Only from the third layer onwards, the reflectivity lacks features that might be attributed to the Ru substrate. For the thicker films (upper panel of Fig. 5) the largest peaks in the reflectivity (peaks marked III and IV, at 10 and 16 eV respectively) do not change their energy with coverage anymore. On the other hand, other smaller peaks clearly change energy when changing the coverage. These latter peaks actually shift in energy in different ways depending on their energy range. Peaks below 10 eV (I and II) shift to higher energy with increasing coverage. For the peak above 16 eV (peak IV), it shifts instead to lower energy.

Although maxima (minima) of the reflectivity are related to a low (high) unocupied density of states, to properly interpret the electron reflectivity data requires not only knowledge of the unoccupied band structure but a multiple scattering calculation. Regular LEED codes are not generally appropriate for the energy range discussed here [19], although in some cases they have been successfully applied [23]. We thus refrain from discussing in more detail the initial stages of the Mg reflectivity. For thicker films, the band structure should correspond closely to the bulk-Mg band structure. We suggest that the broad peaks (III and IV) that do not shift in energy with coverage correspond to gaps in the Mg band structure. The additional oscillations in the reflectivity (I,II, and V) are attributed to quantum interference peaks (QIP) [18] arising from a Fabry-Perot interference effect between electron reflection at the substrate/film and at film/vacuum interfaces [22]. The presence of the vacuum barrier and the interface with the substrate in a thin film of Mg collapse the unoccupied energy bands of bulk Mg into a discrete set of allowed energy values. Unlike for occupied states where quantum well states might appear, for unoccupied states we have quantum well resonances (QWR): they form a discrete set of energies at which electrons can be injected in the Mg film which forms the resonator [20] giving rise to quantum interference peaks. When the number of layers of the film is increased, more allowed QWR appear. To accommodate more energy states in a given bulk band as the thickness is increased, the allowed energy levels shift towards the nearest band edge [20]. The shift in energy in the QIPs as the coverage is increased marked in the top graph of Figure 5 is then naturally explained: for the QIPs marked I and II, the bulk band approaches the nearest band edge at 10 eV. For the QIP marked with IV, the corresponding bulk band presumably approaches the band edge at 16 eV. We note that the presence of these QIPs indicate that the Mg/Ru interface is sharp and devoid of alloying.

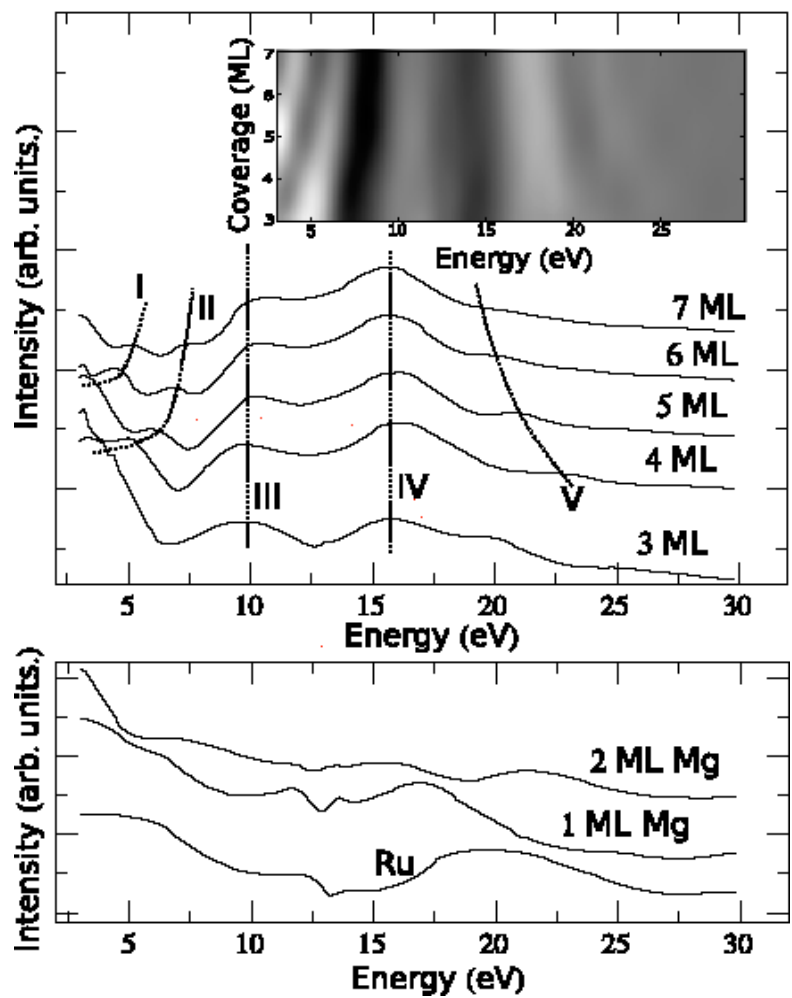

Figure 5: Electron reflectivity of Mg/Ru(0001), shown as a function of energy and coverage. Lower graph: reflectivity for Ru, 1 ML Mg and 2 ML of Mg/Ru(0001). Upper graph: reflectivity for 3-7 ML Mg/Ru(0001). The inset is a bidimensional image showing the change of reflected intensity as a function of both coverage and energy. The data has been differentiated to enhance the contrast. The substrate

temperature was 358 K. The lines show the maxima reflected intensities as a guide to the eye.

## 3.4 Stacking faults in LEEM

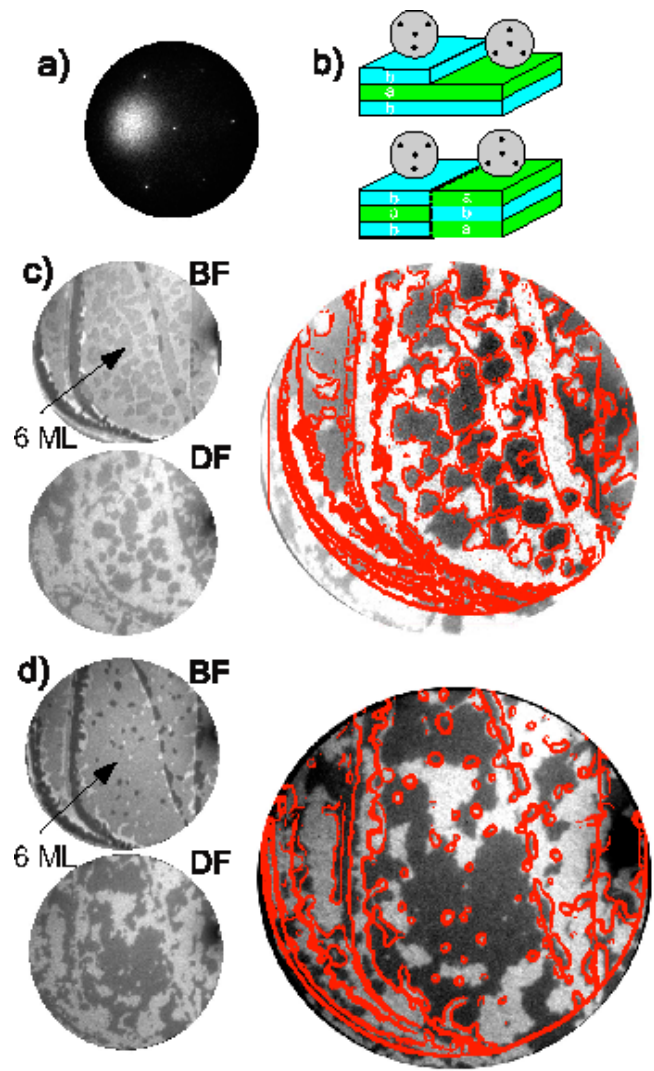

Figure 6: (color online) a) Selected-area diffraction pattern of a 6 ML thick Mg film on Ru(0001) acquired from a region of uniform stacking and thickness. b) Schematic showing how the diffraction patterns should change because consecutive thicknesses have different surface terminations (top) or because layers of the same thickness have different initial stacking (bottom). c) LEEM images of a mostly 5 ML

film with 6 ML thick islands. The field of view is 7 μm. Top: bright field image. bottom: dark field image, reflecting the stacking termination, right: composite image showing the Mg step edges from the bright field image (in red) superimposed on the dark-field image. Electron energies for the bright and dark field images are 3 and 30 eV respectively. d) Same combination of images acquired on a different film (albeit grown on the same substrate terrace) with a nearly complete 6 ML layer. Electron energies for the bright and dark field images are 4.3 and 30 eV respectively.

The stacking sequence of Mg/Ru has been reported as an ...ABAB hexagonal close packed sequence since the beginning of the film growth[5]. (This naming scheme describes the stacking sequence by labeling each possible hexagonal layer in a close packed sequence as A, B or C, with the topmost layer as the rightmost letter.) Nevertheless, there is the possibility of a significant fraction of stacking faults within the films which might arise from the interface or appear later during the film growth. To check for the presence of different surface terminations, dark-field images were acquired in addition to the regular bright-field images. While the latter are created from the specular beam, the dark-field images were formed from one of the first-order diffraction beams (i.e., a (01)-type beam) of the Mg film. As discussed elsewhere[17, 24,25], the three-fold symmetry of the LEED pattern [see in Figure 6(a) the pattern acquired in a film area of uniform thickness and stacking] reflects the symmetry of a given hcp stacking-sequence termination. When changing the stacking termination from ...ABAB to ...BABA, the diffraction pattern rotates by 180º[17]. This can happen when moving across monoatomic steps of the film (we note that the unit cell of an hcp crystal is composed of two hexagonal layers, so consecutive terraces are not crystallographically equivalent), as shown in the top schematic of Figure 6(b). Thus, the dark-field image should show reversed contrast on different surface terminations[17]. This effect is easily observed in a film with intermediate coverage between full layers, such as the film shown in Figure 6(c). An image from the film using the specular beam, i.e., a bright field image, is shown as the first image of Figure 6(c) (marked BF), showing the topography of the film surface, with the different thicknesses presenting different contrast due to the quantum size effect on the reflectivity, as seen in Figures 1 and 2. Selecting one of the first-order diffracted beams by means of an aperture, a dark-field image is shown in Figure 6(c) (marked with DF). In the composite image the islands are outlined in red. At most island boundaries, where the film thickness changes by one layer, there is a bright-dark contrast change in the dark-field image indicating the change in surface termination. Nevertheless, there are some islands that do not show the expected change. The simplest explanation is that the stacking sequence is changed in those islands. In order to check this effect more clearly, a nearly complete-layer film was grown in the same substrate terrace in Figure 6(d). Again the bright field image (marked BF) shows the film topography. The dark field image, as before, reflects the different surface terminations. It is clear from the composite image that even at the same film thickness there are bright and dark areas, which we interpret using the bottom schematic of Figure 6(b). Thus, we find that on a given Ru terrace, the Mg films occurs with two different stackings and the minority stacking covers a significant fraction of the area. The stacking faults might be present from the Mg/Ru interlayer, or they might arise later in the film growth. Further work will be required to determine their location.

## 3.5 Growth in STM: Initial Stages

In addition to using LEEM to study the Mg/Ru growth, we have also employed STM to characterize the growth at room temperature (RT). A first goal was to determine any possible special features of the growth for the first monolayer, where LEEM shows a continuous decrease in reflected intensity up to 0.3 ML, with no clear monolayer island growth detected. This effect was attributed to the growth of Mg islands smaller than the LEEM spatial resolution (see Figure 1b). The STM image in Figure 7a, 4000 Å wide, shows a Mg coverage of less than a complete monolayer, where approximately half of the surface is covered by ramified islands. The gray inset shows an image with the same size from the LEEM (i.e., 4000 Å wide, where the two wide lines correspond to the substrate terraces). As expected, the ramified monolayer islands are below the resolution of our LEEM instrument, and could well be the origin of the "rough" LEEM images of the Mg monolayer. The height of the monolayer islands changes strongly with the tunneling bias. Such effect is clearly seen in the comparison of the STM image acquired with positive and negative bias in Figure 7a (we note that in the STM image the substrate has double steps, 4.2 Å high).

A more detailed image of the Mg film is shown in Figure 7b. The Mg islands have an arrangement of hexagonal protrusions on top of them, up to 0.7 Å high. We interpret these protrusions as reflecting the coincidence pattern between the Mg and the underlying Ru substrate, giving rise to a moiré in real space and to the observed satellites in the LEED pattern. If so, one question is why the pattern is detected only in some parts of the islands (see Figure 7c). Actually, the pattern can be detected also in the areas of the islands that at first appear flat, albeit with a much smaller corrugation, at the resolution limit of our STM electronics. We propose that they actually correspond to molecules on the surface adsorbed at specific positions of the moiré pattern. In any case, the moiré spacing itself is 14 Å. This in-plane periodicity is close to 6 Ru atoms and would correspond to slightly over 5 Mg atoms if their lattice spacing was 3.10 Å. Presumably, this is the origin of the pseudo(5x5) periodicity measured in LEED when the deposited magnesium did not cover the entire Ru surface (see Figure 4).

There seems to be material deposited on the Ru in between the flat Mg islands. An image of these islands with higher resolution shows grain-like structure with a mean size of the grains of around 12 Å (see Figure 7b), albeit with a different orientation than the protrusions on the extended Mg islands. While we do not have an unambiguous explanation, we suggest that these grains are small islands of magnesium surrounded by adsorbed molecules of CO that inhibit further growth. To try to avoid preadsorbed molecules on the surface, the substrate was flashed just before dosing magnesium. Still, the time delay between flashing and dosing would allow the adsorption of a fraction of a monolayer of CO or HO (a few minutes at $5\times10$ mbar), which would be enough to decorate the observed islands. The first islands would then trap the impurities on the surface, and eventually, new magnesium islands would encounter a clean surface and would grow to give rise to the large flat areas. The observed effect is also probably at work in the LEEM images, where the first layers appear more "rough" than later ones.

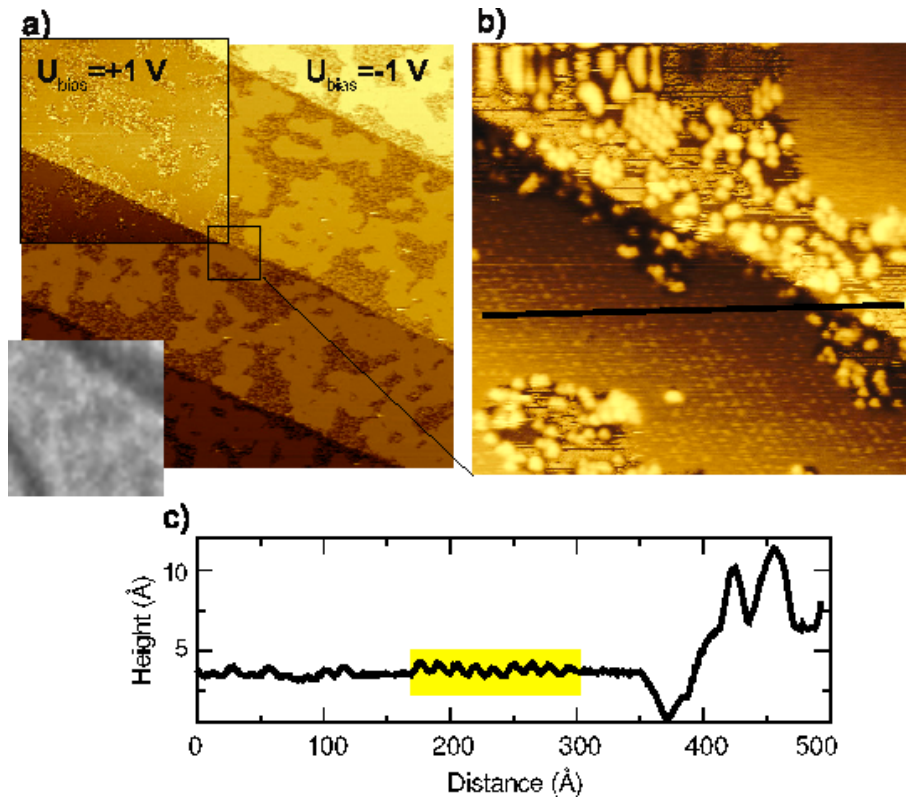

Figure 7: (color online) a) STM image of the first Mg ML evaporated at room temperature over Ru(0001). The image is 4000 Å wide, and the tunneling parameters are $I$=1.27 nA and $U$=-1.06 V. The gray inset shows a section of a LEEM image of the same width (400 nm), with two diagonal substrate steps after growing nearly one complete Mg layer. On the upper top-left corner the same surface is imaged with the opposite bias ($U$=+1.06 V). b) Image 500 Å wide of the same area showing both the hexagonal pattern on the flat islands as well as the features on the uncovered Ru substrate. A double-height substrate step runs diagonally across the image. c) Profile along the line marked in image b). The ordered protrusions on the film are highlighted in yellow.

Dosing more Mg eventually nucleates the second Mg layer. Figure 8 shows a typical image of the substrate totally covered by a monolayer of magnesium and several second layer islands. In the underlying first layer there are several holes extending down to the substrate. These holes are more frequent in the terrace edges and display an irregular shape. The 2 ML islands nucleate preferentially at the lower edge of the substrate steps. The islands show compact edges along the compact directions of the substrate. The LEEM images also detected the preferential nucleation at the step edges (see Figures 1 and 2). But this technique does not directly distinguish which side of the step has the lower or higher terrace, unlike STM.

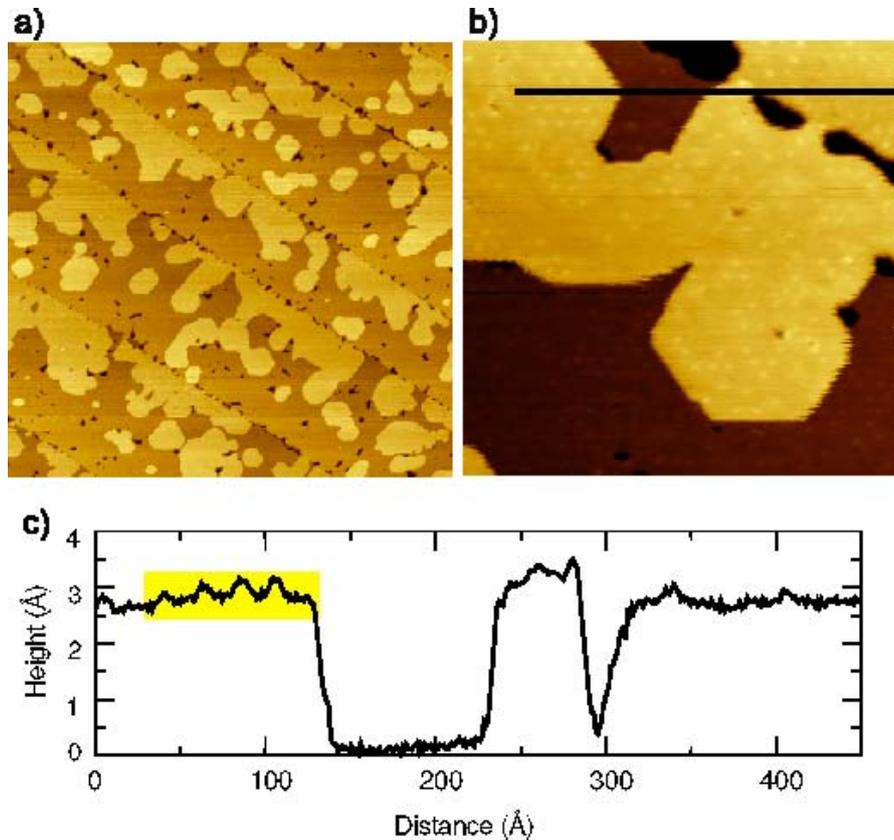

Figure 8: (color online) a) Film with 2 ML islands and a nearly complete first layer. The image is 4000 Å wide. b) Detail of the islands (the image is 500 Å wide). c) Profile along the line marked in image b). The ordered protrusions on the film are highlighted in yellow.

The height of bulk-Ru and bulk-Mg steps are, respectively, 2.11 and 2.60 Å. LEED found Mg layers in a 9 ML Mg film to be separated by 2.64±0.02 Å[4], while the separations between the topmost layers of a monolayer and a two layer films are 2.33±0.04 Å and 2.73±0.03 Å, respectively[5]. The height we obtain relative to the Ru substrate for the submonolayer islands depends strongly on the tunneling bias, with a ×2 difference (from 2Å to 4Å), indicating that is it dominated by electronic effects. On the other hand, the step height for the second monolayer is not strongly bias dependant, and is close to the reported value, 2.6 Å, as show in the profile of Figure 8c. Furthermore, the same profile shows that the step difference between a 2 ML Mg island and a 1 ML Mg region sitting on the adjacent (lower) substrate terrace is 0.5 Å. This corresponds to the difference in step height between Ru and Mg.

As in the submonolayer islands, there are protrusions arranged in an hexagonal pattern on the two layer Mg islands. Their height is 0.4 Å. Their in-plane spacing is 22 Å (shown in the profile presented in Figure 8c), nearly double the spacing of the protrusions for the sub-monolayer Mg islands. The same periodicity is observed on top of the 1 ML and the 2 ML Mg areas. Again, in addition to the clearly defined protrusions we detect a weak corrugation with the same periodicity in otherwise empty areas. We thus assume that the protrusions arise from molecules adsorbed on a moiré pattern. The periodicity of the pattern, 22 Å, agrees with the LEED measurements of the 7x7 compact layer, 21.2±0.5 Å (see Figure 4d). After the Ru surface is completely covered with magnesium, additional Mg densifies the layer, and the in-plane spacing contracts from 3.45 Å to 3.10 Å and the moiré increases smoothly from 12 to 22 Å. The second Mg layer islands start to nucleate only when the latter value is reached. Mg is a system where the moiré periodicity can be continuously tuned between 10 and 20 Å, similar to a few metal/metal systems as Pd/W(100). This behavior might be used to produce ordered arrays of molecules with a tunable separation.

The next layers grow in the same way as the second layer, as shown in Figure 9: the islands nucleate again preferentially at the terrace edges, adjacent to ascending Ru steps. Smaller islands (width of around 200 Å) display sharp edges and hexagonal shapes. We observed no signs of a moiré pattern on Mg thicker than two layers.

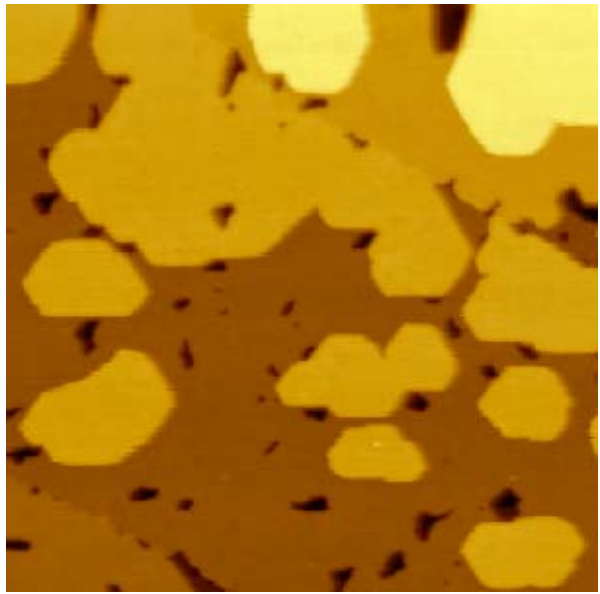

Figure 9: (color online) STM image of 2 ML of Mg with 3 ML islands over Ru(0001). The image is 1500 Å wide.

## 3.6 Growth in STM: Thicker Films

Thicker films contain a high density of points where one or more Mg steps originate, as shown in Figure 10a. These points are where dislocations with a component of their Burgers vector perpendicular to the surface emerge at the film's surface. At the origin of the dislocations we find the same type of sharp steps observed in thinner films and also smooth steps where the height change takes place along a much larger in-plane distance, up to 10 nm (see the profile presented in Figure 10c). Since such screw dislocations can grow new layers without having to nucleate new layers [26], regions with dislocations grow faster. As we do not observe spiral growth in large flat terraces of the substrate, like those followed by LEEM in Figure 2, substrate steps seem to be required to generate the dislocations, together with a minimum film thickness.

Similar observations have been reported in other hcp epitaxial systems, such as Dy films on W(110)[27] or ice on Pt(111)[28]. The same behavior is also described in misoriented nanocrystals[29]. While in the latter case the meeting point of three nanocrystals is the proposed origin of the screw dislocations, in thin films the presumed explanation lies in the coalescence of a minimum of three islands. Two islands are on the same substrate terrace. One of them has a stacking fault. The third island is on an adjacent substrate terrace. We propose, following Ref.[28] that a crucial point in the ultra-thin film case is the presence of stacking faults in the film and substrate steps that differ in height from film steps. (Both can be understood to provide the small misorientation of the nanocrystal case). This kind of interface defect has been much discussed under the general name of a "disconnection"[30]. In the ruthenium substrate, double steps are often observed (see Figure 7a). A Ru double step has a height of 4.2 Å. On the other hand, two Mg (bulk-like) steps are 5.2 Å high. That implies that to connect between them, the Mg layers on both sides of such substrate step are offset by 1.0 Å if they connect islands with the same stacking (see left side islands of the schematics of Figure 10b), or by 2.6-1.0=1.6 Å for islands of different stacking where the first island is one layer thicker (see right side islands of the same schematic). If both types of regions then coalesce, the meeting point will have a screw dislocation, with an abrupt step on one side and a smooth step on the other side of the screw dislocation. These steps initially follow the location of the substrate step. (The order in which the islands connect in Figure 10 has been chosen for clarity.) We believe that this mechanism, or related ones, can explain the presence of screw dislocations in Mg films grown on stepped areas of the Ru substrate as well as their absence in large flat terraces.

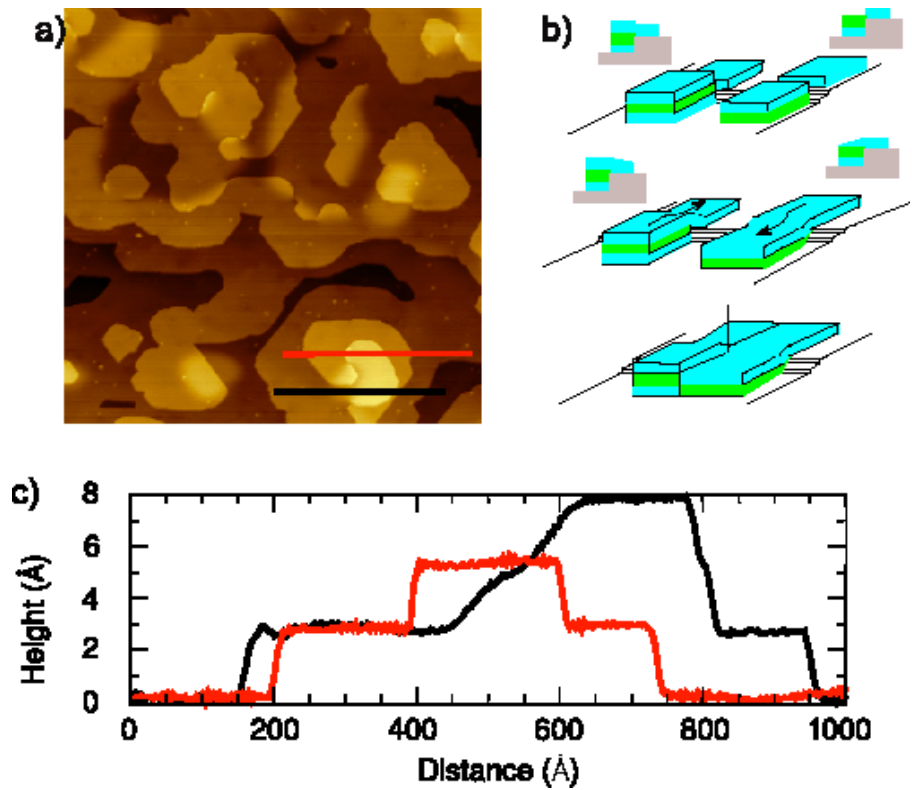

Figure 10: (color online) a) STM image of 5 ML of Mg over Ru(0001). The image size is 2070 Å wide. b) Schematic of a suggested origin of the screw dislocations in the film. On the left, two islands with the same stacking connect across a substrate double step. On the right, the two islands across the Ru double step have different stacking. When the two different connected islands coalesce, they originate a screw dislocation. c) Line profile along the red and black lines in image a).

## 4 Summary

By means of low energy electron microscopy and scanning tunneling microscopy we have characterized in real-space the growth of Mg films on Ru(0001), from less than one layer to 10 ML. The multilayer films, grown at temperatures between room temperature and 390 K, present a strict layer-by-layer growth mode with at most three different levels exposed. The submonolayer islands of Mg on Ru present a moiré pattern with a periodicity of 1.2 nm corresponding to a pseudo-5x5 LEED pattern and consistent with a Mg in-plane lattice spacing of 3.45±0.07 Å. When the monolayer is completed at the start of second layer growth, the layer is compacted reaching a lattice spacing of 3.10±0.06 Å reflected in a real-space moiré pattern of 22 Å and a 7x7 LEED pattern. The moiré pattern is only detected by STM up to the second monolayer of Mg. Thicker films have stacking faults as detected by dark-field LEEM and, on stepped areas, present screw dislocations. The latter arises due to the mismatch in Mg and Ru step heights on stepped areas. Electron reflectivity shows quantum size effects in the unoccupied bands, indicating an abrupt Mg-Ru interface for the thicker films.

## Acknowledgments

This research was supported by the Office of Basic Energy Sciences, Division of Materials and Engineering Sciences, U. S. Department of Energy under Contract No. DE-AC04-94AL85000, and by the Spanish Ministry of Science and Innovation through Project No. MAT2009-14578-C03-01.